# Multiscale phase dynamics & $2\pi$ phase kinks in injection-locked optoelectronic oscillators with large delay

Abhijit Banerjee[1] & Trevor J. Hall[2]

[1]Academy of Technology, Department of Electronics & Communications Engineering, Adisaptagram, Hooghly, West Bengal, 712121, India.

[2]University of Ottawa, Electronic Engineering & Computer Science, Advanced Research Complex, 25 Templeton St, Ottawa, Ontario, K1N 6N5, Canada.

## Abstract

Injection locking of optoelectronic oscillators (OEOs) with large feedback delay gives rise to phase dynamics that lie beyond the scope of classical single-mode locking theory, including the spontaneous formation of persistent 2π phase kinks. In this work, a multiscale theoretical framework is developed that explains the origin, structure, and stability of these phase-slip phenomena in injection-locked OEOs operating in the large-delay regime. Starting from a complex-envelope delay-differential equation that explicitly incorporates hard-limiting gain saturation and RF bandpass filter dynamics, a reduced phase-only description valid for nearly constant oscillation amplitude is derived. Exploiting the separation between fast round-trip dynamics and slow inter-round-trip evolution, a two-timescale reduction yields a continuum of Adler-type equations governing the phase difference between the injected signal and the oscillator as a function of round-trip time. Analytical solutions obtained for weak injection and small detuning show how initially smooth phase profiles sharpen into localized $2\pi$ phase kinks, with $p$ kinks appearing per round trip when the injection is tuned near the $p^{th}$ adjacent mode. Time-domain simulations of the full complex-envelope model validate the predicted phase-kink formation mechanism and reveal the essential role of RF resonator dynamics in determining their persistence. While the reduced phase-only model correctly predicts kink sharpening, resonator-induced amplitude excursions associated with steep phase gradients can erase the kinks when the complex-envelope trajectory fails to encircle the origin, restoring conventional phase locking. These results provide a unified physical interpretation of $2\pi$ phase kinks in injection-locked OEOs and delineate the limits of phase-only models in the presence of large, fast phase transients, identifying a regime of frequency locking without phase locking in large-delay oscillators.

# Injection locked OEO multiscale phase dynamics

## I. Introduction

Optoelectronic oscillators (OEOs) are widely used as low phase-noise radio-frequency (RF) signal sources in advanced communication and radar systems, as local oscillators in frequency synthesis and RF transceivers, and in high-precision measurement and timing applications. Their appeal arises from the combination of ultra-low phase noise, wide tunability, and the ability to generate coherent RF and optical outputs within a single system. Although integrated OEO implementations are increasingly important, non-integrated architectures employing long optical fibre delay lines remain indispensable for achieving the lowest phase-noise performance. These systems also provide a valuable platform for developing and validating theoretical models that capture the essential physics of large-delay oscillators.

Injection locking is a key technique for controlling the frequency stability, spectral purity, and synchronization properties of OEOs. Classical analyses of injection locking are typically based on single-mode oscillator models and lead to Adler-type phase equations that accurately describe locking behaviour when the oscillator spectrum is narrowly concentrated around a single mode. However, OEOs with long feedback delay inherently support a dense comb of oscillation modes, separated in frequency by the inverse of the round-trip delay. As a result, the assumptions underpinning conventional single-mode injection-locking theory may break down, particularly under conditions of weak injection or small detuning where multiple modes can participate in the dynamics.

A delay-differential equation (DDE) framework for injection-locked OEOs that does not rely on a single-mode assumption was introduced in [1]. That work focused primarily on theoretical development and on the accurate prediction of experimentally observed phase-noise spectra but did not explore the full range of nonlinear dynamical behaviours accessible in the large-delay regime. A complementary SIMULINK™-based simulation framework was later presented in [2], providing a detailed time-domain implementation of the injection-locking model and enabling direct visualization of the oscillator dynamics. These simulations revealed the emergence of spiking behaviour in the oscillator phase—specifically, the formation of sharp $2\pi$ phase transitions—under parameter conditions where classical injection-locking theory would predict smooth phase evolution. While these phase spikes were associated with serrodyne-type solutions of idealized reduced phase models in the absence of RF-filtering, a clear physical explanation for their formation and persistence was not established.

Spiking phenomena in optoelectronic and photonic oscillators have attracted considerable interest in recent years, both as a means of emulating neuronal excitability [3] and as a building block for neuromorphic computing systems [4]. In many of these studies, the spikes

# Injection locked OEO multiscale phase dynamics

are broadband amplitude pulses whose temporal width is set primarily by the system bandwidth. Experimental work by Diakonov et al. [5,6] demonstrated the generation of trains of $2\pi$ phase pulses in a broadband OEO under external injection, with the pulse width inversely proportional to the system bandwidth.

The $2\pi$ phase kinks[1] observed in injection-locked OEOs are of particular interest because they correspond to phase-slip events that occur while the instantaneous frequency of the oscillator remains locally locked to the injected carrier. Outside the narrow transition regions associated with the kinks, the phase difference between the oscillator and the injection is nearly constant, yet the oscillator undergoes a net cycle slip on each round trip through the delay line. When coherently mixed with the injected reference, these abrupt phase transitions can be converted into short, ultra-low-jitter RF intensity pulses, enabling the generation of temporal pulse trains and frequency combs with potential applications in precision timing, metrology, and high-speed signal processing.

The spiking phenomena observed in [2] are fundamentally distinct from broadband spiking behaviour reported in optoelectronic and neuromorphic oscillator systems based on low-Q or broadband filtering architectures. In such systems, spikes manifest as broadband amplitude pulses whose temporal width is primarily determined by the overall electrical bandwidth. In contrast, the structures reported in this work are phase kinks in the complex envelope of a narrowband oscillator incorporating a high-Q RF resonator, with transition widths governed by the resonator memory rather than by broadband filtering. The framework developed here explicitly connects phase-kink dynamics to the underlying complex-envelope trajectory and its interaction with RF resonator memory, providing a physical interpretation that is not accessible within purely phase-only or broadband spiking models.

The present work makes four principal contributions. First, starting from a complex-envelope delay–differential equation that explicitly includes hard-limiting gain saturation and RF resonator dynamics, a reduced phase-only description of injection-locked optoelectronic oscillators is derived that remains valid in the large-delay regime beyond the assumptions of classical single-mode locking theory. Second, by exploiting the natural separation between fast round-trip dynamics and slow inter-round-trip evolution, a two-timescale reduction of the phase-only model yields a continuum of Adler-type equations defined over the round-trip fast time. Third, analytical solutions of these equations reveal a robust physical mechanism for the formation of localized $2\pi$ phase kinks, predicting the appearance of $p$ kinks per round trip when the injected carrier is tuned near the $p^{th}$ adjacent oscillation mode. Fourth, time-domain simulations results of the full complex-envelope model are presented which demonstrate that while the reduced phase-only theory correctly predicts kink formation and sharpening, the persistence or decay of these structures is ultimately governed by RF resonator dynamics, establishing clear limits on the applicability of phase-only models. Together, these results identify and characterise a distinct dynamical regime of injection locking in which frequency locking occurs without global phase locking.

---

[1] Throughout this paper, the term *phase kink* refers exclusively to a localized 2π phase transition in the complex envelope occurring over a small fraction of the round-trip time. The term is not used to describe broadband amplitude spikes or phase pulses arising in low-Q or broadband oscillator architectures.

The remainder of the paper is organised as follows. Section II describes the architecture of a single-loop optoelectronic oscillator under RF injection. Section III develops the complex-envelope model, including the representation of saturated gain and RF resonator dynamics. In Section IV, a reduced phase-only model is derived using the Leeson approximation, and Section V introduces the two-timescale analysis leading to a continuum of Adler equations and their analytical solutions. Section VI presents simulation results that validate the theory and delineate its domain of applicability. Finally, Section VII summarizes the key findings and discusses their implications for the modeling of injection-locked OEOs and related large-delay oscillators.

## II. Optoelectronic oscillator architecture

The single-loop optoelectronic oscillator (OEO) shown schematically in Figure 1 comprises an RF-photonic section and an RF-electronic section connected in a feedback loop. The RF-photonic section includes a laser (LD), a Mach–Zehnder modulator (MZM), a single-mode optical-fibre (SMF) delay line (FDL), and a photodetector (PD). The RF-electronic section consists of an electronic amplifier (EA) chain that incorporates an RF resonator, which functions as a bandpass filter (EBPF). The PD recovers the RF modulation, which is then amplified by the RF electronic section and used to drive the MZM, thereby closing the loop. An external RF signal can also be injected via a second input port of the electronic coupler (EC). The OEO is an instance of a time delay oscillator (TDO) where the RF-photonic section serves as a long *RF delay line*, and the RF electronic section serves as a *sustaining RF amplifier*.

Under small-signal conditions, both the RF amplifier chain and the MZM behave linearly. At higher drive levels, however, their intrinsic nonlinearities generate harmonic distortion. The EBPF suppresses these unwanted harmonics so that the fundamental component alone experiences large-signal gain saturation. Although the MZM can be the sole source of gain saturation, it is preferable to limit the saturated RF output such that the MZM drive spans only the interval between adjacent transmission minima and maxima but no further. This approach maximises modulation depth while avoiding oscillation-envelope instabilities at elevated MZM drive powers [7].

# Injection locked OEO multiscale phase dynamics

Together, the EBPF and FDL form a compound resonator. By exploiting the exceptionally low loss of optical fibre, the FDL introduces a long delay $\tau_D$ that is essential for achieving the low phase-noise levels characteristic of high-performance OEOs. The EBPF supplies the selectivity needed to favour single-mode oscillation and contributes a smaller delay $\tau_R$ to the total loop round-trip time.

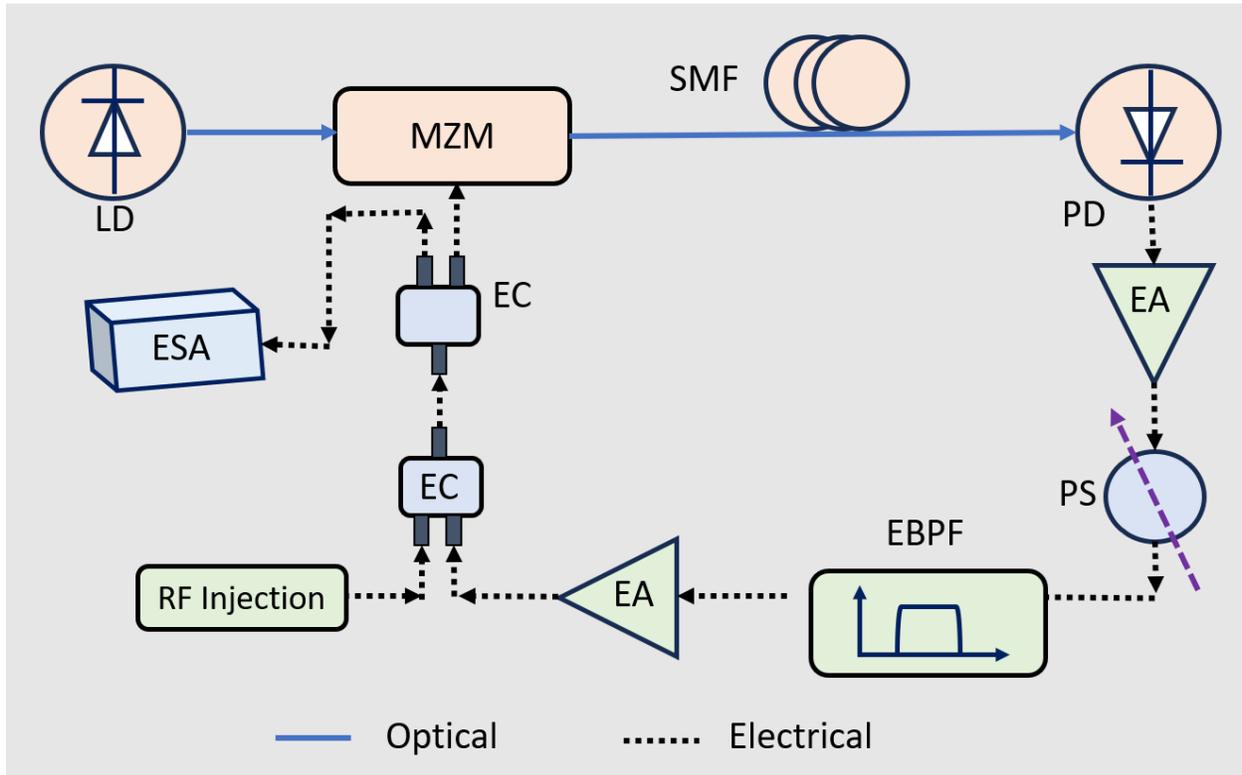

Figure 1 Schematic of a single-loop OEO under injection by an RF source. LD: semiconductor laser diode; MZM: Mach-Zehnder Modulator; SMF: single mode optical fibre delay line; PD: photodetector, EA: electronic amplifier (RF amplifier); PS: phase shifter (tuning); EBPF: electronic bandpass filter (BPF); EC: electronic coupler; ESA: electronic spectrum analyser.

# Injection locked OEO multiscale phase dynamics

## III. Complex-envelope model

### Model formulation

The dynamics of a time-delay oscillator subject to external injection are described using a complex-envelope representation of the fundamental harmonic. Let $u$ denote the complex envelope following the point of injection, $v$ the envelope prior to saturated gain, and $\bar{v}$ the envelope after saturated gain. The injection (forcing) term is denoted by $w$. The model is specified by the relations:

$$u = \bar{v} + w \quad ; \quad \bar{v} = e^{i\phi_v}\kappa(|v|)v \quad ; \quad v = h \otimes (D_{\tau_D}u)$$

*Equation 1*

where $\kappa$ is the saturated (large-signal) gain, $h$ is the impulse response of the baseband equivalent of the RF resonator, $\otimes$ denotes convolution, $D_{\tau_D}$ is the delay operator with time delay $\tau_D$, and $\phi_v$ collects the phase shift contributions of the loop components. The standard assumption that the saturation mechanism is invariant to the phase has been made. The delay operator is defined by:

$$(D_{\tau_D}f)(t) = f(t - \tau_D)$$

*Equation 2*

with the delay time $\tau_D$ set by the optical fibre delay line within the loop.

Consolidating the relationships of Equation 1 yields a single delay-integral equation:

$$v = h \otimes D_{\tau_D}\left(e^{i\phi_v}\kappa(|v|)v + w\right)$$

*Equation 3*

### RF Resonator representation

The RF resonator is modeled at baseband by a single pole low-pass filter frequency response:

$$H(\omega) = \frac{1}{1 + i\omega\tau_R}$$

*Equation 4*

which is associated with the differential equation:

$$\tau_R \frac{dv}{dt} + v = u$$

*Equation 5*

where $u$, $v$ are respectively the complex envelope at the input and output ports of the RF resonator and $\tau_R$ is a characteristic time constant.

The corresponding impulse response is:

# Injection locked OEO multiscale phase dynamics

$$h(t) = \begin{cases} 0 & ; \quad t < 0 \\ \dfrac{1}{\tau_R}\exp\left(-\dfrac{t}{\tau_R}\right) & ; \quad t \geq 0 \end{cases}$$

*Equation 6*

This standard reduction captures the dominant resonator dynamics while enabling a compact envelope-domain description.

## Saturated-Gain Laws

### Hard-limiting amplifier

The saturated gain $\kappa$ is real-valued and decreases from the small-signal (linear) gain $\gamma$ in such a manner that the magnitude of the amplifier output remains approximately constant when the amplifier operates in saturation. For a hard-limiting amplifier driven by a locally pure carrier $a\cos(t)$, the large-signal gain is given by the expression [2]:

$$\kappa(|a|) = \gamma \begin{cases} 1 & ; \quad |a| \leq 1 \\ 1 - \dfrac{1}{\pi}(2\theta - \sin(2\theta)) & , \quad \theta = \cos^{-1}(1/|a|) \quad ; \quad |a| > 1 \end{cases}$$

*Equation 7*

which encodes the fraction of the cycle spent in limiting through the parameter $\theta$.

### Mach–Zehnder Modulator (MZM) biased at quadrature

For a Mach–Zehnder modulator biased at quadrature, the (intensity) transfer function is sinusoidal up to a constant that is removed by subsequent filtering. The modulation recovered by the photodiode can be represented via:

$$v(t) = \sin(a\cos(t))$$

*Equation 8*

and, invoking the Jacobi–Anger expansion, the corresponding large-signal gain is:

$$\kappa(|a|) = \gamma \frac{J_1(|a|)}{|a|}$$

*Equation 9*

where $J_1$ is the first order Bessel function of the first kind. These expressions furnish a convenient parametric description of the fundamental component under large-signal drive.

## Consolidated delay–differential equation description

Equation 5 may be used to reduce the Equation 3 to a first order delay–differential equation (DDE) for the complex envelope $v$:

$$\tau_R \frac{dv}{dt} + v = D_{\tau_D}\big(\kappa(|v|)e^{i\phi_v}v + w\big)$$

# Injection locked OEO multiscale phase dynamics

*Equation 10*

Equation 10 compactly captures the dispersive memory time $\tau_R$, the explicit time delay $\tau_D$, the nonlinear saturated gain $\kappa$, and the effect of external injection $w$. It provides a tractable basis for analysis of injection locking, multistability, and oscillation onset in time delay oscillators and in particular optoelectronic oscillators.

## IV. Reduced phase-only model

In the absence of forcing ($w = 0$), the trivial solution ($v = 0$) of Equation 10 is unstable if the net linear gain exceeds unity ($\gamma > 1$). Then, initiated by noise, multimode oscillation rapidly grows in the transient linear regime until the combined saturation of the sustaining amplifier and the Mach–Zehnder modulator (MZM) suppresses further growth in the established oscillation regime. Under certain restrictions, the oscillation magnitude $|\bar{v}|$ may be assumed constant[2]. It is then convenient to introduce the polar representation of the complex envelopes:

$$u = a_u \exp(i\theta_u) \quad ; \quad \bar{v} = \bar{a}_v \exp(i\theta_v) \quad ; \quad w = a_w \exp(i\theta_w)$$

*Equation 11*

where the magnitudes $a_u, \bar{a}_v, a_w$ and phases $\theta_u, \theta_v, \theta_w$ are real variables.

Hence:

$$u = \bar{v} + w \quad \Rightarrow \quad r \exp(i\psi) = 1 + \rho \exp(i\theta)$$

*Equation 12*

where:

$$r = a_u/\bar{a}_v, \quad \rho = a_w/\bar{a}_v, \quad \psi = \theta_u - \theta_v, \quad \theta = \theta_w - \theta_v$$

*Equation 13*

$r$ is the injection-induced gain/loss, $\rho$ is the injection ratio, $\psi$ is the injection-induced phase shift, and $\theta$ is the phase difference between the complex envelope of the injection $w$ and the oscillation $v$. Presuming $\rho$ is known, equating real and imaginary parts of Equation 12 enables the elimination of one of the three variables $r, \psi, \theta$. Eliminating $r$ yields:

$$\psi = \tan^{-1}\left(\frac{\rho \sin(\theta)}{1 + \rho \cos(\theta)}\right)$$

*Equation 14*

---

[2] The assumption holds if $\kappa(|a|)|a|$ is a bounded monotonic increasing function of $|a|$, which is satisfied for sustaining amplifier dominated saturation, or the net linear gain $\gamma$ is lower than a threshold for an envelope modulation instability, which for MZM-dominated saturation requires $1 < \gamma < 2.316$.

# Injection locked OEO multiscale phase dynamics

where the arctangent should be interpreted in the four-quadrant sense:

$$\psi = \arg(x + iy) \quad ; \quad x = 1 + \rho\cos(\theta) \quad , \quad y = \rho\sin(\theta)$$

*Equation 15*

To complete the reduction to a phase-only model, a phase-only model of the RF resonator is required. The Leeson approximation [8,9] assumes that the differential operator describing the action of the RF resonator on the complex envelope also applies to its action on the phase of the complex envelope:

$$\tau_R \frac{dv}{dt} + v = u \quad \Rightarrow \quad \tau_R \frac{d\theta_v}{dt} + \theta_v = \theta_u$$

*Equation 16*

This approximation is valid for slow phase fluctuations provided their frequency fluctuations are small $|d\theta_u/dt| \ll 1$. It is also valid for small but fast phase fluctuations $|\theta_u|\ll 1$. These two constraints are satisfied by oscillators with spectral components concentrated well within the passband apart from a small largely white noise component. Consequently, the Leeson approximation is ubiquitous in its application to oscillators because of its simplicity, linearity and utility.

The Leeson approximation reduces the complex envelope model Equation 10 to a first order delay–differential equation (DDE) in the phase only:

$$\tau_R \frac{d\theta_v}{dt} + \theta_v = D_{\tau_D}(\phi_v + \psi + \theta_v)$$

*Equation 17*

where equality in equations involving phase-only representations of complex envelopes must be interpreted modulo $2\pi$.

In the supplementary material it is shown that:

$$\tau_R \frac{dv}{dt} + v = u \quad \Rightarrow \quad v - \frac{1}{2}\tau_R^2 \frac{d^2v}{dt^2} + \cdots = D_{\tau_R} u$$

*Equation 18*

which reveals that the primary effect of the RF resonator is to introduce an additional delay equal to the on-resonance group delay $\tau_R$. The second order derivative accounts for the very slow evolution of an established multimode oscillation state of the free oscillator towards a single-mode oscillation state.

The evolution of the oscillation state following the onset of injection for practical parameter values is rapid albeit still slow on the scale of $\tau_D$ so the second order derivative may be neglected reducing Equation 17 to:

# Injection locked OEO multiscale phase dynamics

$$\theta_v = D_{\tau_G}(\phi_v + \psi + \theta_v) \quad ; \quad \tau_G = \tau_D + \tau_R$$

*Equation 19*

Since $\psi$ depends only on the phase difference $\theta = \theta_w - \theta_v$ it is convenient introduce a tuning phase shift parameter $\phi_w$ defined by:

$$\theta_w = D_{\tau_G}(\phi_w + \theta_w)$$

*Equation 20*

If the oscillator is in a single-mode oscillation state with frequency $\omega_V$ before the onset of injection ($\psi(\theta) = 0$) and $\omega_w$ is the frequency of the injected carrier then:

$$\theta_v = \omega_V t \quad ; \quad \theta_w = \omega_w t$$
$$\Rightarrow$$
$$\phi_v = \omega_V \tau_G \quad ; \quad \phi_w = \omega_w \tau_G$$
$$\Rightarrow$$
$$\phi = \phi_w - \phi_v = (\omega_w - \omega_V)\tau_G$$

*Equation 21*

Assuming the spectrum of oscillation and injection are concentrated well within the passband $\omega \tau_R \ll 1$ of the RF resonator, Equation 21 remains valid for multimode oscillation and injection where $\omega_w, \omega_V$ are then parameters that specify the detuning of the oscillation and injection frequency combs from the regular frequency grid $\{\omega_p | \omega_p \tau_G = 2p\pi \; p \in \mathbb{Z}\}$ of the free oscillator for no detuning $\phi_v = 0$. This enables the treatment of the special case where the injection is tuned to a neighbourhood of an adjacent sidemode to the principal oscillating mode prior to the onset of injection.

The reduced phase-only model of injection locking of a time delay oscillator with large delay follows from subtraction of Equation 19 from Equation 20:

$$\theta = D_{\tau_G}(\phi - \psi(\theta) + \theta) \quad ; \quad \psi(\theta) = \tan^{-1}\left(\frac{\rho \sin(\theta)}{1 + \rho \cos(\theta)}\right) \quad ; \quad \theta = \theta_w - \theta_v$$

*Equation 22*

Equation 22 captures established multimode oscillation while treating the phase as the sole state variable. It is a difference equation analogue for injection locking of a time delay oscillator of the Paciorek [10] variant of the Adler differential equation [11] for injection locking of a classical oscillator.

The Leeson approximation employed to obtain the reduced phase-only model assumes either small, fast phase fluctuations or slow phase variations whose associated instantaneous frequency excursions remain well within the passband of the RF resonator. While these conditions are appropriate for steady-state phase-noise analysis and for smoothly varying phase dynamics, they are not uniformly satisfied in the presence of large,

# Injection locked OEO multiscale phase dynamics

fast phase transients such as the $2\pi$ phase kinks considered in this work. Steep phase gradients can induce significant amplitude modulation at the output of the RF resonator, violating the constant-amplitude assumption underlying the phase-only description. Consequently, the reduced model should be interpreted as predictive of phase-kink formation and sharpening mechanisms, but not necessarily of their long-term persistence.

This section has established a reduced phase-only description of injection locking in large-delay oscillators, clarifying both its analytical tractability and its inherent limitations. This formulation serves as the foundation for the two-timescale analysis developed in the next section, which exposes dynamical behaviour inaccessible within single-mode or instantaneous phase-locking models.

## V. Two-timescale analysis

In the large-delay regime, the established oscillation may be viewed as an almost-periodic function of time, with fast dynamics occurring on the scale of the round-trip group delay $\tau_G$ and slow evolution unfolding over many round trips. The two-timescale representation introduced here formalises this structure by embedding the phase dynamics on a cylindrical domain, with the fast time parameterizing position along a round-trip and the slow time governing inter-round-trip evolution. Each point along the fast-time coordinate obeys an Adler-type evolution equation, but these equations are not dynamically independent: they are coupled implicitly through the initial phase profile and through the action of RF resonator filtering, which enforces continuity and smoothness over a finite memory interval $\tau_R$. This interpretation is essential for understanding how initially smooth phase profiles can evolve into sharply localised, but not strictly discontinuous, phase kinks.

For large delay, simulations demonstrate that the complex envelope of an established oscillation is almost periodic with period $\tau_G = \tau_D + \tau_R$ with the waveform within each period slowly evolving with time. The asymptotic analysis of the characteristic equation of the linear free oscillator presented in the Supplementary Material provides theoretical support for this intuition. This motivates introducing variables $\tilde{f}$ that are functions of two times $(t_1, t_2)$ that are $\tau_G$-periodic in $t_1$ and vary slowly in $t_2$ and which agree with the original variable $f$ on $t_1 = t_2 = t$:

$$f(t) = \tilde{f}(t,t) \quad ; \quad \tilde{f}(t_1 + \tau_G, t_2) = \tilde{f}(t_1, t_2)$$
$$\Rightarrow$$
$$D_{\tau_G}\tilde{f} \approx \tilde{f} - \tau_G \frac{\partial \tilde{f}}{\partial t_2} + \cdots$$

*Equation 23*

This two-time structure can be visualized by winding $f(t)$ around a circular cylinder with circumference $\tau_G$, with the fast time $t_1$ as the angular coordinate , and the slow time $t_2$ as

# Injection locked OEO multiscale phase dynamics

the axial coordinate, which with suitable interpolation creates a function $\tilde{f}$ defined everywhere on the surface of the cylinder.

The difference equation (Equation 22) thereby may be approximated up to first order in $\tau_G$ by the *partial differential equation:*

$$\tau_G \frac{\partial \tilde{\theta}}{\partial t_2} + \tan^{-1}\left(\frac{\rho \sin(\tilde{\theta})}{1 + \rho \cos(\tilde{\theta})}\right) = \phi$$

*Equation 24*

This equation provides solutions in $t_2$ stemming from each point $t_1$ on the boundary defined by $t_2 = 0$. The initial condition $\tilde{\theta}(t_1, 0)$ is a $\tau_G$-periodic function modulo $2\pi$. Detuning of the oscillator and injection from the regular frequency grid $\{\omega_p | \omega_p \tau_G = 2p\pi \; p \in \mathbb{Z}\}$ is accounted for by the evolution in $t_2$ of $\tilde{\theta}_v$, $\tilde{\theta}_w$ through the detuning parameters $\phi_v$, $\phi_w$. To preserve the assumed slow variation in $t_2$ requires small detuning which is consistent with the small lock-in range for weak injection $\rho \ll 1$. Under these conditions Equation 24 reduces to a continuum of Alder equations:

$$\tau_G \frac{\partial \tilde{\theta}}{\partial t_2} + \rho \sin(\tilde{\theta}) = \phi$$

*Equation 25*

one for each $t_1$, that admit analytic solutions detailed in the Supplementary Material.

The analytic solutions evolve according to functions with argument $\gamma t_2$ or $\varpi t_2$ where:

$$\gamma = \frac{1}{2}\frac{1}{\tau_D}\sqrt{\rho^2 - \phi^2} \quad ; \quad \varpi = \frac{1}{2}\frac{1}{\tau_D}\sqrt{\phi^2 - \rho^2}$$

*Equation 26*

Slow variation in $t_2$ requires

$$\gamma, \varpi \sim O(\varepsilon) \quad \Rightarrow \quad \phi, \rho \sim O(\varepsilon)$$

*Equation 27*

Confirming that the validity of the multiscale solutions requires weak injection and small detuning.

The analytic solutions are continuous in $t_2$. Nevertheless, even if the initial condition function $\tilde{\theta}(t_1, 0)$ is mod $2\pi$ continuous in $t_1$ the value of neighboring points in $t_1$ of $\tilde{\theta}(t_1, t_2)$ can depart from each other as $t_2$ advances, introducing singularities. In practice, the RF resonator will smooth the complex envelope over an effective interval of $\tau_R$ preserving continuity in $t_1$. However, $\tau_R \ll \tau_D$ so, although technically continuous solutions, singular-like behaviour can develop. This is the origin of $2\pi$ phase kinks observed experimentally [5,6] and in simulation [2].

## Injection locked OEO multiscale phase dynamics

The analytic solutions of the Adler equation relevant to the present problem include both hyperbolic tangent (Equation 59) and hyperbolic cotangent (Equation 57) branches, depending on the initial phase offset relative to the equilibrium point. While the hyperbolic tangent solution is commonly emphasised in standard treatments, the hyperbolic cotangent solution is essential here, as it naturally introduces singular crossings corresponding to phase values equivalent modulo $2\pi$. These singular solutions provide the analytical origin of phase-slip events and are indispensable for describing the evolution of localised $2\pi$ phase kinks within the two-timescale framework.

This section has demonstrated how the interplay between fast round-trip structure and slow inter-round-trip evolution gives rise to localised $2\pi$ phase kinks as natural solutions of the two-timescale Adler equation. These results provide a direct analytical mechanism for phase-slip formation that is subsequently tested and refined using full complex-envelope simulations.

# VI. Results

To isolate the deterministic large-signal dynamics responsible for phase-kink formation and evolution, all simulations reported here are performed in the absence of additive noise and are initialised from a pure single-mode oscillation prior to the onset of injection. This approach significantly reduces computational overhead while preserving the nonlinear phase and amplitude interactions of interest. Stochastic effects such as noise-induced nucleation, timing jitter, and rare switching events are beyond the scope of the present study and will be addressed separately.

The two-timescale Adler equation derived in Section V admits analytical solutions that provide direct insight into the phase dynamics of injection-locked optoelectronic oscillators with large delay. However, the reduced phase-only formulation, in its present form, does not fully capture the dynamical response of the RF resonator beyond its on-resonance group delay. While it is in principle possible to include a Leeson-type model of the RF resonator within the two-timescale framework, the Leeson approximation remains valid only for sufficiently small phase excursions or for large phase variations occurring over timescales long compared to the resonator memory. The $2\pi$ phase kinks of interest here constitute fast, large phase transients that violate these conditions. Consequently, the full complex-envelope model is required to faithfully describe their interaction with the RF resonator.

To assess the domain of validity of the analytical predictions and to investigate the persistence or decay of $2\pi$ phase kinks, time-domain simulations of the complex-envelope model of a single-loop OEO under RF injection are performed using a SIMULINK™-based graphical block-diagram implementation. The simulation methodology and representative subsystem configurations are described in detail in [2]. Here, the simulations are

# Injection locked OEO multiscale phase dynamics

specifically used to: (i) validate the mechanism of phase-kink formation predicted by the two-timescale Adler model; and (ii) elucidate the role of RF resonator dynamics in shaping the observed phase transient behaviour.

In the simulation model, two saturating amplifiers are employed: one preceding the RF resonator to suppress amplitude modulation induced by the injected signal, and one following the RF resonator to remove amplitude modulation arising from the conversion of phase modulation into amplitude modulation by the resonator. This configuration ensures sustained operation in a regime that closely approximates the assumptions underlying the reduced phase-only description with an exact nonlinear phase representation of the RF resonator, insofar as the complex-envelope trajectory does not pass through the origin. To focus on large-signal deterministic behaviour, small-signal noise sources are disabled, thereby reducing simulation time while preserving the dynamics of interest. Prior to the onset of injection, the oscillator state is initialised to a pure single-mode oscillation corresponding to the main mode ($p = 0$). Unless otherwise stated, the simulation parameters are those listed in Table I.

| | |
|---|---|
| $\tau_D = 24.9\ \mu s$ | Delay line delay time |
| $\tau_R = 0.1\ \mu s$ | Resonator on-resonance group delay |
| $\tau_G = \tau_D + \tau_R = 25\ \mu s$ | Round-trip group delay time |
| $\Delta f = 1/\pi\tau_R = 3.18\ MHz$ | Resonator bandwidth |
| $FSR = 1/\tau_G = 40\ kHz$ | Frequency interval between modes |
| $\rho = 0.075, 0.1$ | Injection ratio |
| $p = 0,1,2$ | Mode index |
| $\omega\tau_G = 2p\pi + \phi$ | Principal part ($2p\pi$) and secondary part ($\phi$) of the detuning ($\omega\tau_G$) of the injected carrier. |
| $t_s = 0.01\ \mu s$ | Sample time |

*Table 1 Simulation model parameters*

## Phase-kink formation and comparison with analytical solutions

Figure 2 illustrates the formation of a single $2\pi$ phase kink per round-trip when the injected carrier is tuned close to the first adjacent mode ($p = 1$). Figure 2(a) presents simulation data from the complex-envelope model, rendered as a surface plot of the phase difference $\theta$ between the injected signal and the oscillator over the fast-time/slow-time plane. Figure 2(b) shows the corresponding analytical solution obtained from the two-timescale Adler equation. The qualitative agreement between the two is striking: in both cases, a sharp phase transition develops and persists across successive round trips.

# Injection locked OEO multiscale phase dynamics

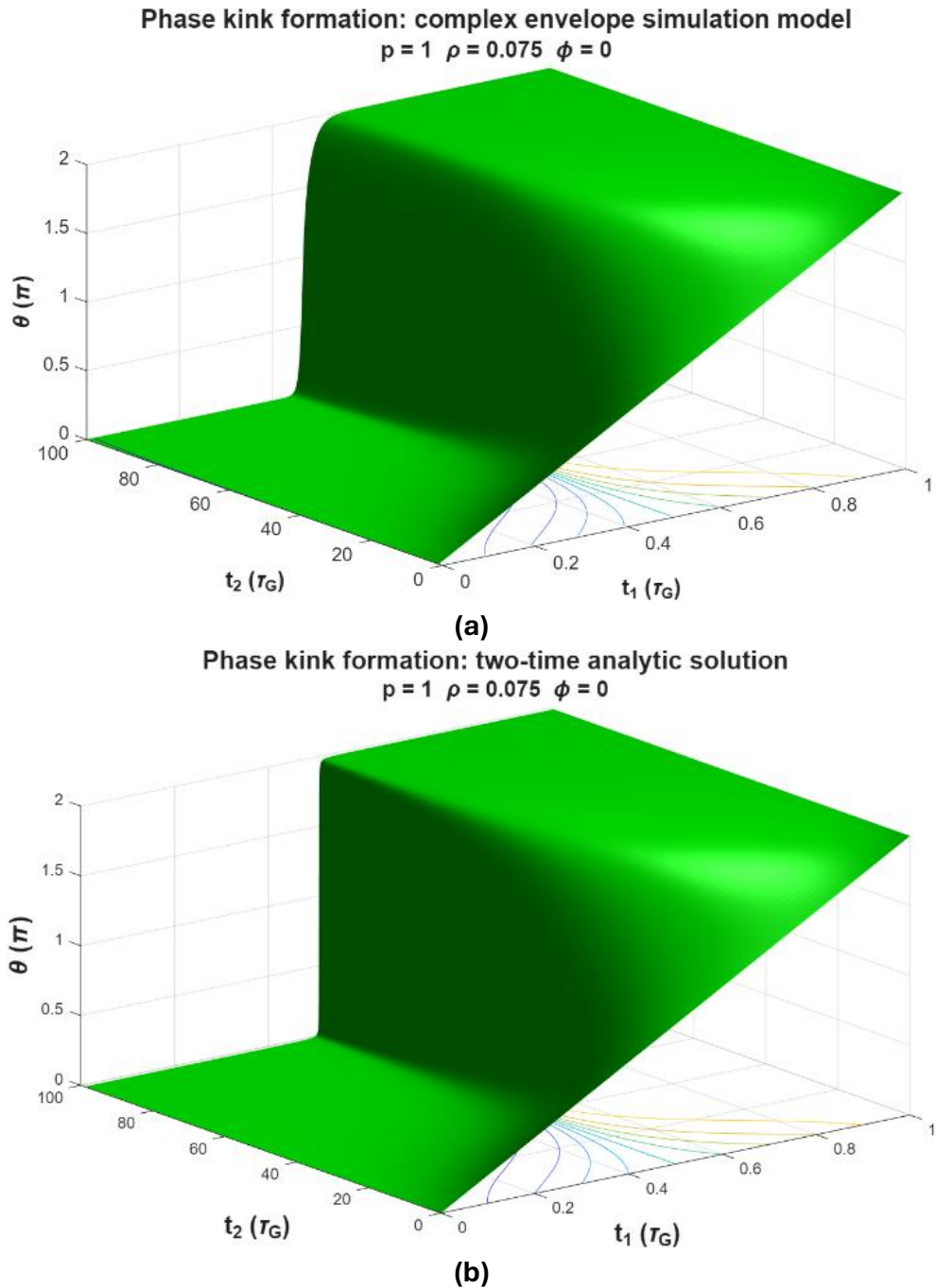

Figure 2. Phase kink formation by an optoelectronic oscillator subject to injection tuned to the first adjacent mode ($p = 1$). The x-axis corresponds to the fast time $t_1$ within one round-trip, while the y-axis represents slow time $t_2$ measured in units of the round-trip group delay $\tau_G$. The z-axis corresponds to the phase difference $\theta$ in units of $\pi$ radians between the injected carrier and the oscillator. Isophase contours of $\theta$ are distinguished by colour. The injection ratio is $\rho = 0.075$ and the detuning from the adjacent mode is $\phi = 0$. (a) Time domain simulation data raster scanned to form a surface plot of $\theta$ over $(t_1, t_2)$ illustrating the formation of a $2\pi$ phase kinks over successive round trips. (b) The two-time analytic solution for comparison.

# Injection locked OEO multiscale phase dynamics

The most significant quantitative difference between the analytical and simulation results is the structure of the transition region itself. In the analytical solution, the phase kink asymptotically approaches an ideal step function, reflecting the absence of any fast-time smoothing mechanism in the reduced model. In contrast, the simulation data exhibit a finite transition width, which is set by the memory of the RF resonator. This behaviour is consistent with the expectation that the resonator smooths phase variations over a timescale comparable to its group delay $\tau_R$.

Figure 3 shows analogous results for injection tuned near the second adjacent mode ($p = 2$). In this case, two $2\pi$ phase kinks are observed within each round-trip interval. Both the complex-envelope simulations and the analytical solutions predict the formation of $p$ phase kinks per round-trip when the injection is tuned near the $p^{th}$-adjacent mode. This systematic correspondence provides strong evidence that the fundamental mechanism responsible for phase-kink formation is captured by the two-timescale Adler description, despite its neglect of detailed RF resonator dynamics.

## Frequency locking with phase slipping

Outside the narrow transition regions associated with the phase kinks, the phase difference between the injected signal and the oscillator remains approximately constant. In these regions, the instantaneous frequency of the oscillator is therefore locked to the injected carrier frequency. However, within each transition region, the phase slips by approximately $2\pi$, such that the oscillator undergoes a net cycle slip once per phase kink. Consequently, the average phase slope over one round trip remains equal to that of the free-running main mode. The oscillator is thus frequency locked but not phase locked, occupying a distinct dynamical regime that is not described by classical injection-locking theory.

## Persistence versus decay of phase kinks

The analytical solutions of the two-timescale Adler equation predict that phase-kink transition regions tend to narrow from one round trip to the next. In the full complex-envelope simulations, this narrowing is counteracted by the smoothing action of the RF resonator, which broadens the phase transition through its finite memory. The persistence or decay of phase kinks is therefore determined by the balance between these two competing effects.

Figure 4 illustrates this interplay by showing the magnitude of the complex envelope at the output of the RF resonator as a function of time. For relatively weak injection ($\rho = 0.075$), Figure 4(a) shows that downward spikes in the magnitude accompany the narrowing of the phase-kink transition region, but these spikes stabilize at a fixed depth and do not reach zero. In this regime, the broadening induced by the RF resonator compensates the narrowing predicted by the analytical model, resulting in persistent $2\pi$ phase kinks.

# Injection locked OEO multiscale phase dynamics

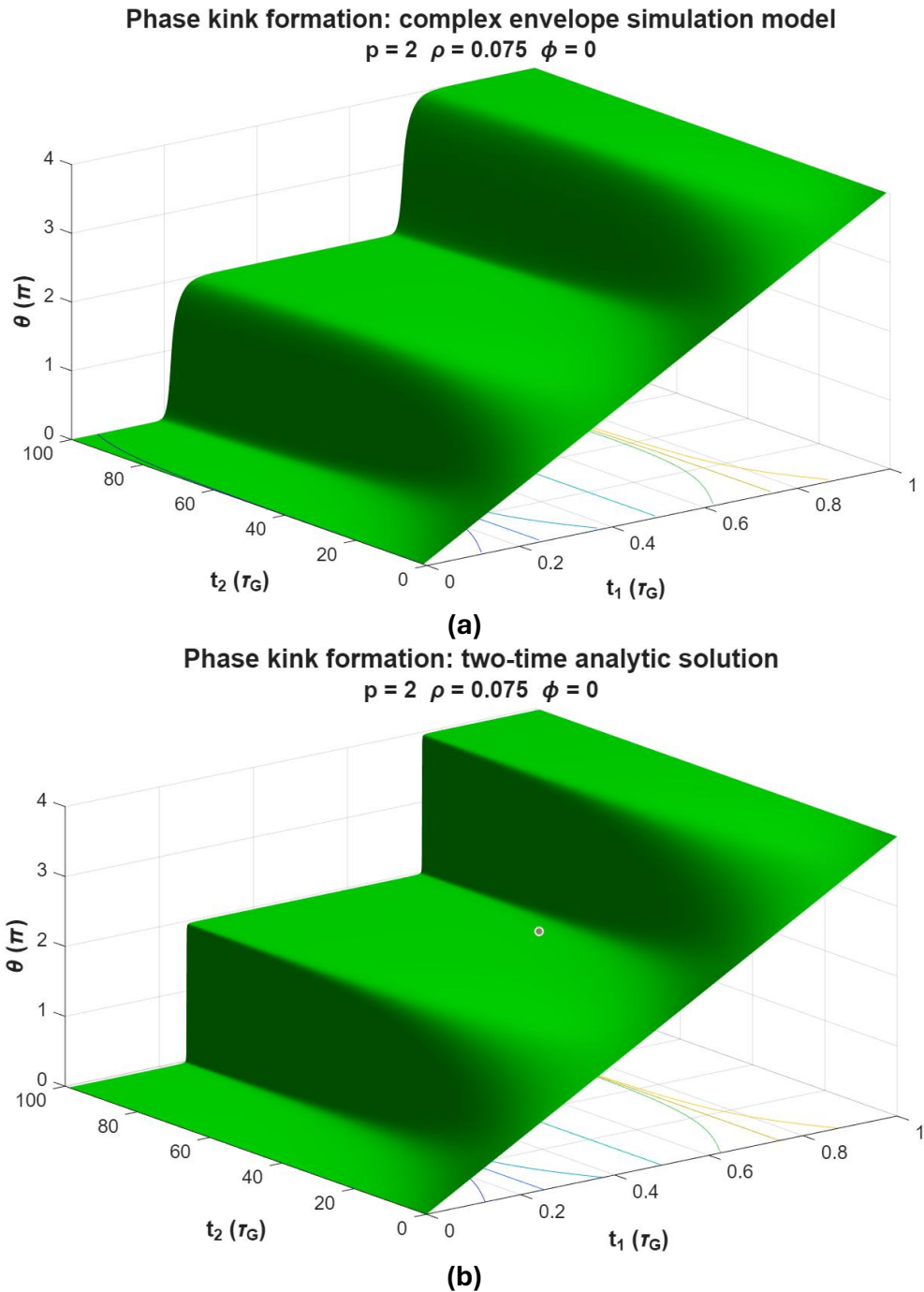

(a)

(b)

*Figure 3. Phase kink formation by an optoelectronic oscillator subject to injection tuned to the second adjacent mode ($p = 2$) with all other parameter values the same as Figure 2. Two $2\pi$ phase kinks are observed within each round-trip time $\tau_D$ (a) Time domain simulation data raster scanned to form a surface plot of $\theta$ over $(t_1, t_2)$. (b) The two-time analytic solution for comparison.*

# Injection locked OEO multiscale phase dynamics

For stronger injection ($\rho = 0.1$), shown in Figure 4(b), the narrowing dominates. As the transition region becomes sufficiently steep, the complex-envelope trajectory at the resonator output approaches and eventually passes through the origin. This event, marked by a magnitude spike reaching zero, heralds rapid decay of the phase kink and subsequent phase locking to the injected carrier.

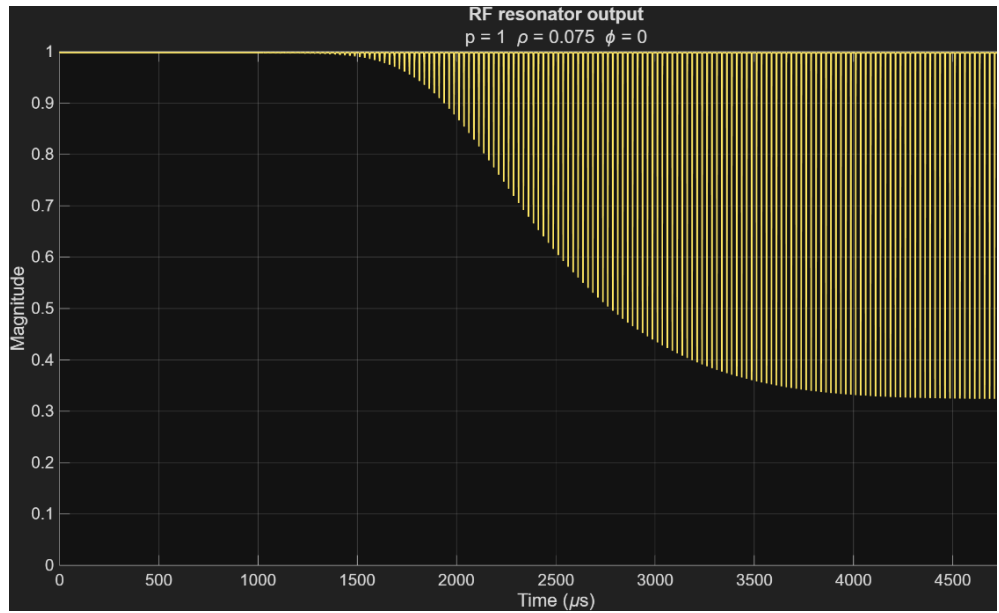

(a)

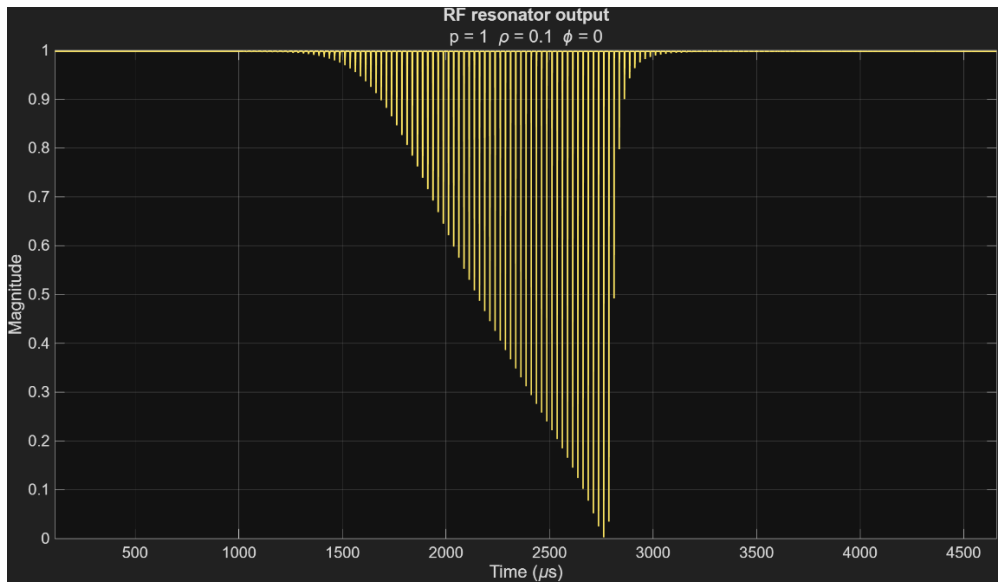

(b)

*Figure 4. The magnitude of the output of the RF resonator versus time in µs. On reach round trip the narrowing of phase kink is opposed by broadening by the RF resonator. (a) For an injection ratio $\rho = 0.075$, the broadening prevails leading to persistent $2\pi$ phase kinks. (b) For an injection ratio $\rho = 0.1$, the narrowing prevails leading to rapid decay of the phase kinks heralded by a magnitude spike reaching zero.*

# Injection locked OEO multiscale phase dynamics

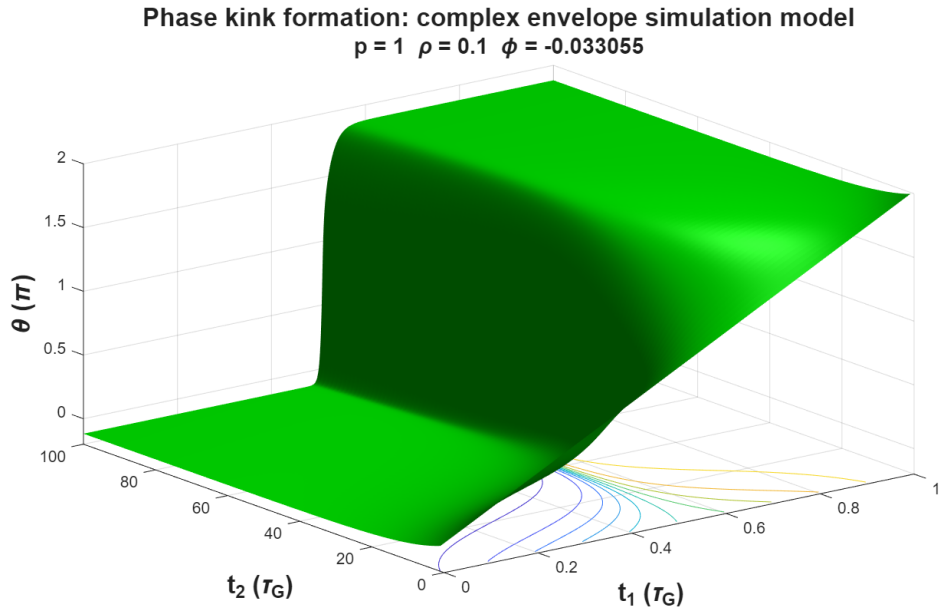

(a)

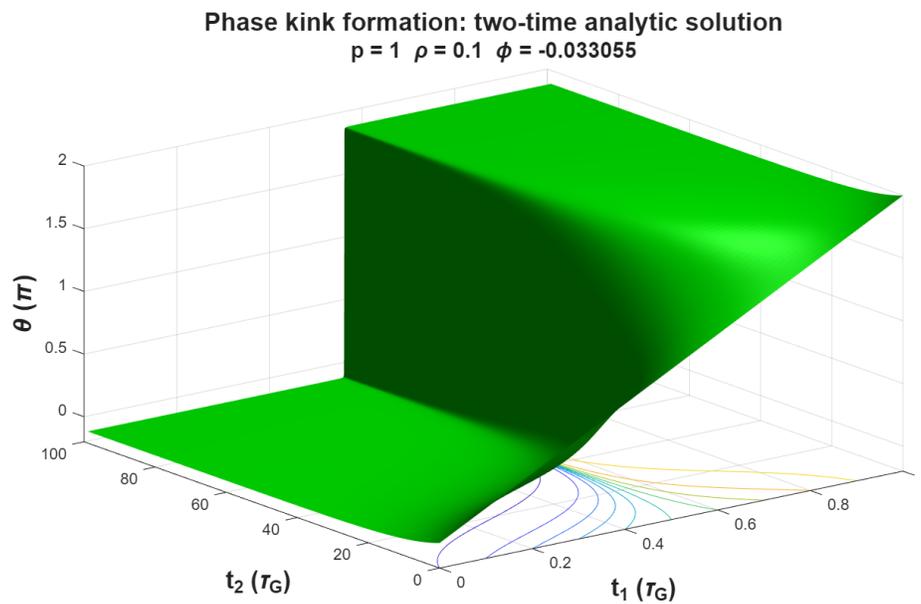

(b)

*Figure 5. Phase-kink formation by an optoelectronic oscillator subject to injection tuned to the first adjacent mode (p = 1). The injection ratio is ρ = 0.1. For zero detuning (ϕ = 0) from the adjacent mode the 2π phase kink decays but detuning by ϕ = −0.033 restores a persistent phase kink. All other parameters are the same as Figure 2. (a) Complex envelope model simulation data. (b) The two-time analytic solution for comparison.*

# Injection locked OEO multiscale phase dynamics

Figure 5 demonstrates that non-zero detuning from the adjacent mode can restore persistent phase kinks in cases where they would otherwise decay. While the growth rates appearing in the analytical solutions depend on the difference of squares between the injection ratio $\rho$ and the secondary detuning parameter $\phi$, the simulations reveal a more nuanced dependence on detuning. Persistence is not invariant to the sign of $\phi$ alone but is invariant under reversal of the total detuning, including both the principal and secondary components. Phase-kink persistence is favoured when the injected carrier is detuned closer to the main mode, an intuitively reasonable result that is only partially captured by the reduced phase-only model.

To further elucidate the critical role of the RF resonator dynamics in the $2\pi$ phase kink decay process, a SIMULINK™ simulation was created illustrated in Figure 6 that synthesises a $\tau_D$-periodic continuous phase $\theta_u$ that shares the principal qualitative features of the observed $2\pi$ phase kink parameterised by a single steepness parameter $\eta$. After conversion to a unimodular complex envelope $u$ it is applied to the complex-envelope model of the RF resonator. The input $u$ and output $v$ along with their associated unwrapped phases $\theta_u$ and $\theta_v$ are collected for different values of $\eta$. The model is described by:

$$\tau_R \frac{dv}{dt} + v = u = \exp(i\theta_u) \quad ; \quad \theta_u(t) = \pi \tanh\big(\eta(\tau_D/\pi)\tan((\pi/\tau_D)(t - \tau_D/2))\big)$$

*Equation 28*

The data is visualised after sufficient time for the initial transient response of the RF resonator to have decayed to negligible proportions. This is aided by consistent initial conditions for the RF resonator $v(0) = -1$ and the phase $\theta_u(0) = -\pi$. The steepness parameter $\eta$ is a measure of the peak instantaneous frequency. It is equal to the phase slope at the midpoint ($t = \tau_D/2 \ mod \ (\tau_D)$) of the synthesised waveform measured in units of $\pi$ per unit time.

The trajectory of the complex envelope on the complex plane is illustrated in Figure 7(a) for a representative set of steepness parameter values $\eta = 5, 12.89, 12.90, 15 \ (\pi/\mu s)$. For $\eta \ll 5 \ (\pi/\mu s)$ the trajectory in the complex plane follows the trajectory of the input (blue) lagging by the group delay $\tau_R$. For $\eta = 5 \ (\pi/\mu s)$ the trajectory is distorted in shape because of the induced variation in the magnitude of the RF resonator output $v$. Nevertheless, Figure 7 (b) demonstrates the phase of the output essentially still follows the input phase with the same lag. For $\eta = 12.89 \ (\pi/\mu s)$ the trajectory approaches the origin very closely; the downward spike in magnitude thereby almost reaches zero. The close approach to the origin results in a rapid change of phase that results in an almost vertical $\sim \pi$ phase jump in Figure 7 (b) (solid red line). Nevertheless, the trajectory still encircles the origin preserving the overall $2\pi$ phase increment within each period.

# Injection locked OEO multiscale phase dynamics

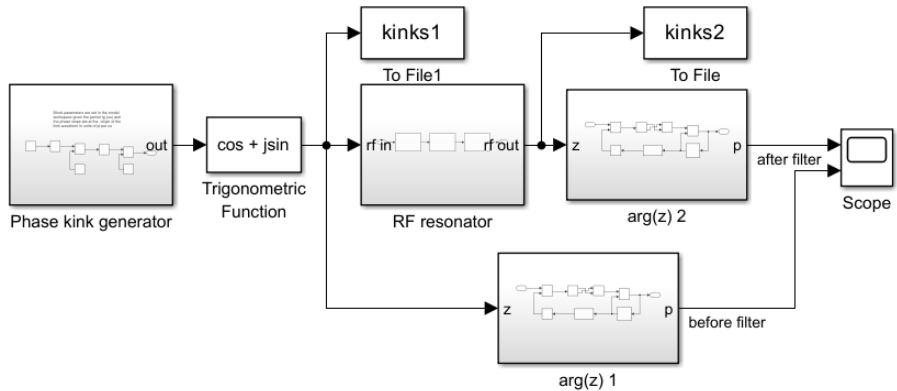

(a) RF resonator driven by $2\pi$ phase kink train

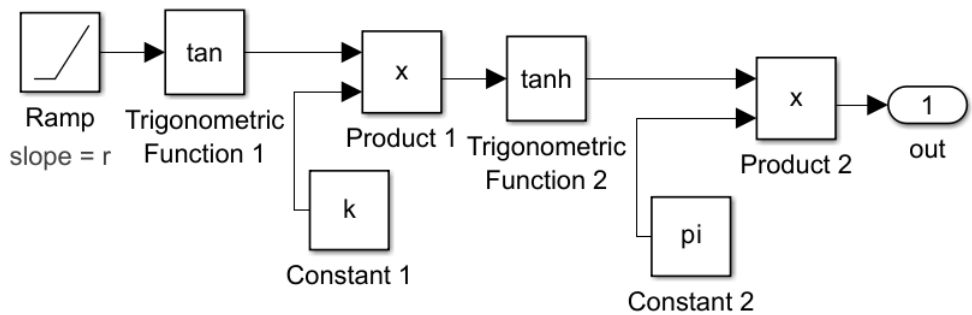

(b) Phase kink generator subsystem

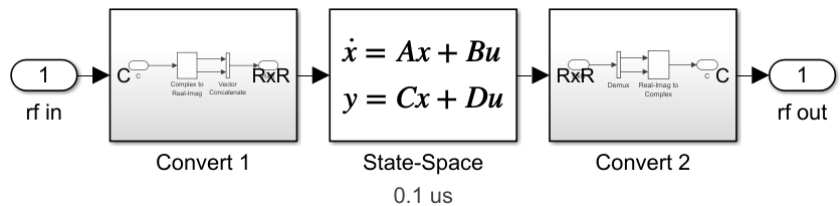

(c) RF resonator subsystem

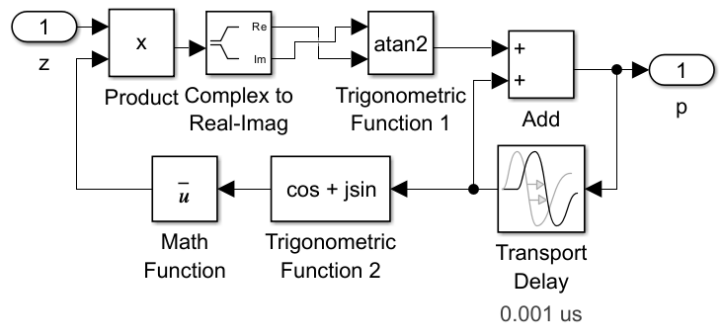

(d) arg(z) subsystem

*Figure 6 Simulink™ model to study the behavior of the RF resonator's response to a narrow $2\pi$ phase pulse. (a) Test harness: the Scope block measures the time evolution of the unwrapped phase before and after the filter, the To file blocks record the complex envelope associated with the phase before and after the filter. (b)&(c) expansion of Phase kink generator and RF resonator subsystems. (d) expansion of arg(z) subsystem that extracts the unwrapped phase from the complex envelope.*

# Injection locked OEO multiscale phase dynamics

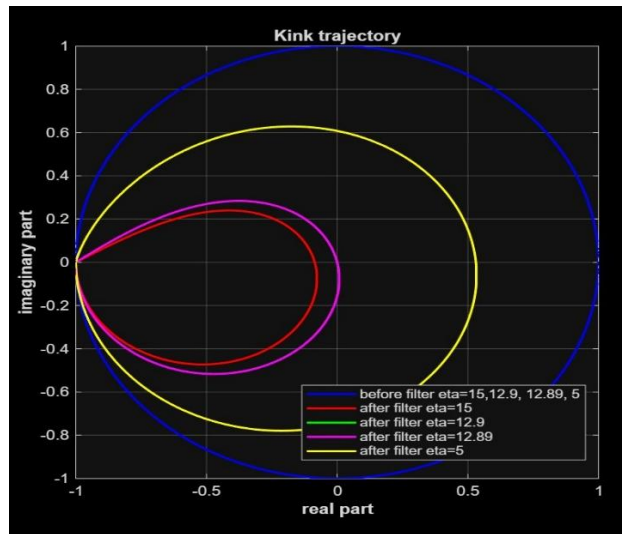

(a) Trajectories on the complex plane of the complex envelope before and after filtering

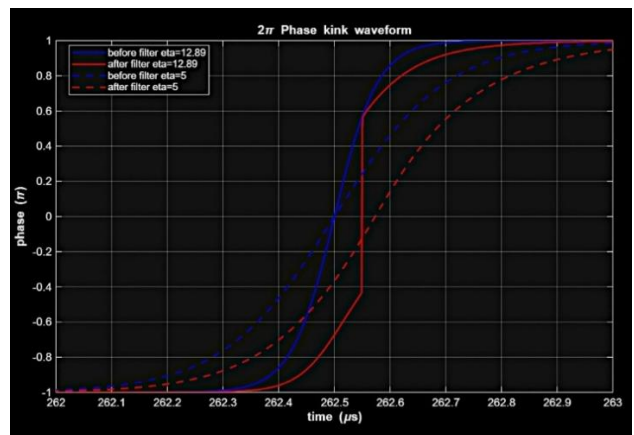

(b) Time domain plots phase of the complex envelope before and after filtering for trajectories encircling the origin in (a).

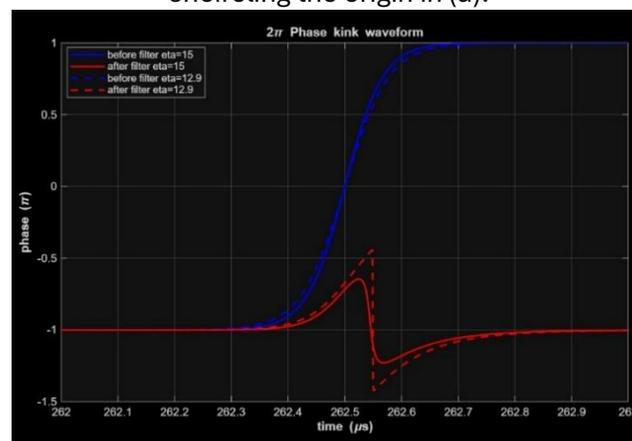

(c) Time domain plots phase of the complex envelope before and after filtering for trajectories not encircling the origin in (a).

*Figure 7 Simulation results of the action of the RF resonator on $2\pi$ phase kinks for steepness parameter values $\eta = 5, 12.89, 12.90, 15$ ($\pi/\mu s$). The period is $\tau_G = 25\ \mu s$ and the on-resonance group delay is $\tau_R = 0.1\ \mu s$.*

# Injection locked OEO multiscale phase dynamics

For $\eta = 12.90$ $(\pi/\mu s)$ the trajectory bypasses the origin resulting in a dramatic change to the behaviour of the time domain phase as seen in Figure 7 (c). The close approach to the origin still results in an almost vertical $\sim \pi$ phase jump but now it has the opposite sign. Since the trajectory no longer encircles the origin the overall phase increment with each period is now zero with a maximum excursion of $\pm \pi/2$. The transition of behaviour triggers the rapid erasure of the phase kink by the smoothing provided by the resonator. The $\eta = 15$ $(\pi/\mu s)$ case confirms that the total phase increment is zero and the steep negative phase jump is smooth and not an artefact provided the trajectory does not pass exactly through the origin.

The persistence or decay of a $2\pi$ phase kink is thereby closely linked to the topology of the complex-envelope trajectory at the output of the RF resonator. When the trajectory encircles the origin of the complex plane once per round trip, the associated phase undergoes a net $2\pi$ increment and the phase kink is preserved. If, however, steep phase gradients drive the trajectory sufficiently close to or through the origin, the encirclement is lost and the total phase excursion becomes bounded by $\pm \pi/2$, leading to rapid erosion of the kink and restoration of conventional phase locking. This origin-encirclement criterion provides a physically transparent explanation for the limits of phase-kink persistence in realistic oscillators.

Although the numerical results presented here employ a representative set of parameters corresponding to a practical fibre-based OEO, the qualitative phenomena reported—namely phase-kink formation, the appearance of $p$ kinks when injection is applied near the $p^{th}$ adjacent mode, and the competition between kink sharpening and resonator-induced smoothing—are generic features of large-delay oscillators incorporating narrowband RF filtering. The specific persistence thresholds depend quantitatively on ratios such as $\tau_D/\tau_R$ and on the injection strength $\rho$ and detuning $\phi$, but the underlying mechanisms do not rely on fine parameter tuning.

# VII. Conclusions

This work has presented an investigation of the phase dynamics of injection-locked optoelectronic oscillators with large feedback delay with particular emphasis on the origin, persistence, and decay of $2\pi$ phase kinks. A reduced phase-only description applicable in the large-delay regime has been derived starting from a complex-envelope delay–differential equation that explicitly incorporates hard-limiting gain saturation and RF bandpass filter dynamics. A two-timescale analysis of this reduced model yields a continuum of Adler-type equations governing the phase difference between the injected signal and the oscillator.

Analytical solutions of the resulting two-timescale Adler equation reveal a clear physical mechanism for phase-kink formation: initially smooth phase profiles evolve into increasingly sharp transitions due to slow inter-round-trip dynamics acting on fast round-

# Injection locked OEO multiscale phase dynamics

trip phase structure. The analysis predicts the formation of $2p\pi$ phase kinks per round trip when the injection is tuned near the $p^{th}$. adjacent mode, as well as the narrowing of the transition regions over successive round trips.

Time-domain simulations of the full complex-envelope model validate these analytical predictions and demonstrate that the essential mechanism of phase-kink formation is accurately captured by the two-timescale phase-only framework. The simulations also reveal the critical role played by RF resonator dynamics in determining the ultimate fate of the phase kinks. Transient amplitude excursions induced by steep phase gradients, which are absent from the reduced phase-only description, may cause the complex-envelope trajectory to pass through the origin, leading to rapid erosion of the phase kink and restoration of conventional phase locking. The balance between analytical narrowing and resonator-induced smoothing therefore determines whether phase kinks persist or decay.

These results provide a unified physical interpretation of $2\pi$ phase kinks in injection-locked OEOs and clarify the limits of reduced phase-only models when applied to large, fast phase transients. More broadly, the analysis highlights how multiscale phase dynamics in large-delay oscillators can give rise to non-classical injection-locking behaviour, including regimes of frequency locking without phase locking. The framework developed here is expected to be applicable to a wider class of delayed oscillatory systems and may prove useful for the controlled generation and manipulation of phase-encoded pulse sequences in photonic and microwave systems.

## Acknowledgements

The authors are grateful to Dr. Mehedi Hasan for his continuing engagement in our research and specifically for creating Figure 1. Dr. Trevor Hall is grateful to the University of Ottawa for their support of his University Research Chair.

## References:

[1 M. Hasan, A. Banerjee, T. J. Hall, 'Injection locking of optoelectronic oscillators with large delay', J. Lightwave Technology, **40**(9), 2754–2762, (2022).

[2] A. Banerjee, T. J. Hall, 'Simulation of optoelectronic oscillator injection locking, pulling and spiking phenomena', Scientific Reports, **14**, 4332, (2024).

[3] D. P. Rosin, 'Pulse train solutions in a time-delayed opto-electronic oscillator', Master's thesis, Technische Universität Berlin, (2011).

[4] H. Tian, L. Zhang, Z. Zeng, Y. Wu, Z. Fu, W. Lyu, Y. Zhang, Z. Zhang, S. Zhang, H. Li, H. Y. Liu, 'Externally triggered spiking and spiking cluster oscillation in broadband optoelectronic oscillator', J. Lightwave Technology, **41**(1), 48-58, (2022).

# Injection locked OEO multiscale phase dynamics

[5] A. Diakonov, M. Horowitz, 'Generation of ultra-low jitter radio frequency phase pulses by a phase-locked oscillator', Opt. Lett., **46**(19),5047-5050, (2021).

[6] V. Smulakovsky, A. Diakonov, A. Katzenlson, M. Horowitz, 'Temporal locking of pulses in injection locked oscillators,' Scientific Reports, **15**, 5602, (2025).

[7] E. C. Levy, M. Horowitz, C. R. Menyuk, 'Modeling optoelectronic oscillators', J. Opt. Soc. Am. B, **26**(1),148-159, (2009).

[8] D. B. Leeson, "A simple model of feedback oscillator noise spectrum," Proc. IEEE, 329-330, (1966).

[9] D. B. Leeson, 'Oscillator phase noise: a 50-years review', IEEE Trans. Ultrasonics, Ferroelectrics & Frequency Control, **63**(8), 1208-1225, (2016).

[10] L. J. Paciorek, 'Injection locking of oscillators', Proc. IEEE, **53**(11), 1723-1727, (1965).

[11] R. Adler, "A study of locking phenomena in oscillators," *Proc. IRE*, 351-357, (1946).

[12] R. M. Corless, G. H. Gonnet, D. E. G. Hare, D. J. Jeffrey, D. E. Knuth, 'On the Lambert W function', Advances in Computational Mathematics **5**(1) 329-359 (1996).

# Injection locked OEO multiscale phase dynamics

## Supplementary Material

### Linear analysis of the free oscillator

The complex envelope DDE model (main text Equation 10) is linearized by treating the gain $\kappa$ as a constant independent of the oscillation magnitude $|v|$. For simplicity, the case of a free ($w = 0$) oscillator with unit gain ($\kappa = 1$) and zero detuning ($\phi_v = 0$) is first considered so that:

$$\tau_R \frac{dv}{dt} + (v - D_{\tau_D} v) = 0$$

*Equation 29*

The substitution:

$$v = \exp(st) \quad ; \quad s = \sigma + i\omega$$

*Equation 30*

yields the characteristic equation:

$$1 + s\tau_R = \exp(-s\tau_D)$$

*Equation 31*

which may be rearranged into the same form:

$$(1 + s\tau_R)\rho \exp\big((1 + s\tau_R)\rho\big) = \rho \exp(\rho) \quad ; \quad \rho = \tau_D/\tau_R$$

*Equation 32*

as the definition of the Lambert W function:

$$W(z) \exp\big(W(z)\big) = z$$

*Equation 33*

Consequently, Equation 31 admits a countable infinity of roots given by:

$$s\tau_R = -1 + \frac{1}{\rho} W_p(\rho \exp(\rho)) \quad ; \quad p \in \mathbb{Z}$$

*Equation 34*

where $W_p$ is the $p^{\text{th}}$ branch of the complex Lambert function [12]. Each root is associated to a distinct mode of the free oscillator via Equation 30. These roots may be readily calculated numerically to high precision using, for example, the MATLAB™ function lambertw.

Equation 31 may be expressed equivalently by taking its real and imaginary parts to yield the two coupled equations:

$$1 + \rho^{-1}\sigma\tau_D - \exp(-\sigma\tau_D) \cos(\omega\tau_D) = 0$$
$$\rho^{-1}\omega\tau_D + \sin(\omega\tau_D) = 0$$

*Equation 35*

which may be formally solved recursively using the series expansion:

## Injection locked OEO multiscale phase dynamics

$$\sigma = \sigma_0 + \rho^{-1}\sigma_1 + \rho^{-2}\sigma_2 + \cdots$$
$$\omega = \omega_0 + \rho^{-1}\omega_1 + \rho^{-2}\omega_2 + \cdots$$

*Equation 36*

The result up to second order is:

$$\sigma\tau_D = -\frac{1}{2}(\omega\tau_R)^2 \quad ; \quad \omega(\tau_D + \tau_R) = 2p\pi \quad ; \quad p \in \mathbb{Z}$$

*Equation 37*

which is accurate provided all significant modes have frequencies that fall well within the passband of the RF resonator ($|\omega\tau_R| \ll 1$) (see Figure 8).

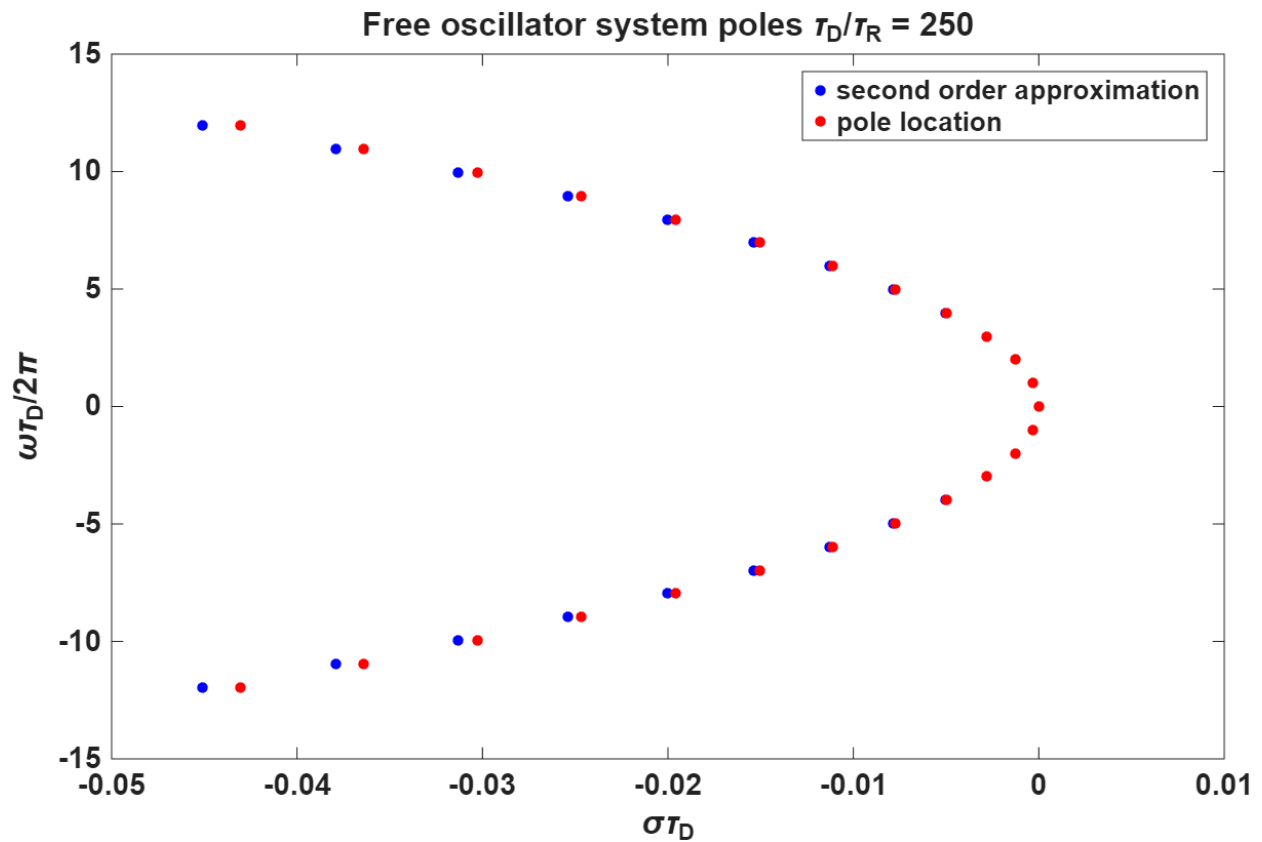

*Figure 8 Comparison of the free oscillator exact pole locations (red) and their second order asymptotic approximation (blue) given by Equation 37 for $\tau_D/\tau_R = 250$. The error is barely visible for the first 11 poles centred on the origin and for the 25 poles potted the resonant frequency (omega) remains close to the exact value. The principal deviation is an overestimate of the damping (sigma) that slowly increases for poles further from the origin.*

# Injection locked OEO multiscale phase dynamics

By linearity, the general solution of the linearized oscillator (Equation 29) is given by the weighted sum:

$$v(t) = \sum_{p=-\infty,\infty} c_p \exp(\omega_p t) \exp(\sigma_p t)$$

*Equation 38*

where $c_p$ is the complex amplitude of mode $p$.

$v$ may be equivalently expressed as:

$$v(t) = \tilde{v}(t,t) \quad ; \quad \tilde{v}(t_1, t_2) = \sum_{p=-\infty,\infty} c_p \exp(\omega_p t_1) \exp(\sigma_p t_2)$$

*Equation 39*

It follows from Equation 37 that $\tilde{v}$ periodic in $t_1$ with period

$$\tilde{v}(t_1 - \tau_G, t_2) = \tilde{v}(t_1 - \tau_G, t_2) \quad ; \quad \tau_G = \tau_D + \tau_R$$

*Equation 40*

and slowly varying in $t_2$ on the timescale of the period provided the oscillation spectrum is concentrated well within the passband of the RF resonator so that: $[\omega_p \tau_R] \ll 1$. Consequently, $v$ is an almost-periodic function that slowly evolves in time as is observed in simulations of established oscillations of the complex-envelope model for large delay and including saturated gain, detuning and injection.

## RF resonator delay & diffusion contributions

Consider the baseband equivalent low-pass filter representation of the RF resonator (main text Equation 8):

$$\tau_R \frac{dv}{dt} + v = u$$

*Equation 41*

Operate on both sides by the delay operator $D_{\tau_R}$ and apply Taylor's theorem:

$$D_{\tau_R} v = v + \sum_{n=1,\infty} \frac{(-1)^n}{n!} \tau_R^n \frac{d^n v}{dt^n} \quad \Rightarrow \quad \tau_R D_{\tau_R} \frac{dv}{dt} = -\sum_{n=1,\infty} \frac{(-1)^n n}{n!} \tau_R^n \frac{d^n v}{dt^n}$$

*Equation 42*

Hence:

$$\tau_R \frac{dD_{\tau_R} v}{dt} + D_{\tau_R} v = v - \sum_{n=2,\infty} (-1)^n (n-1) \frac{1}{n!} \tau_R^n \frac{d^n v}{dt^n} = v - \frac{1}{2}\tau_R^2 \frac{d^2 v}{dt^2} + \frac{1}{3}\tau_R^3 \frac{d^3 v}{dt^3} - \frac{1}{8}\tau_R^4 \frac{d^4 v}{dt^4} + \cdots$$

*Equation 43*

# Injection locked OEO multiscale phase dynamics

which up to second order accuracy simplifies to:

$$v - \frac{1}{2}\tau_R^2 \frac{d^2v}{dt^2} = D_{\tau_R} u$$

*Equation 44*

## General solution of the Adler equation

### Derivation

The Adler equation is given by:

$$\tau \frac{d\theta}{dt} + \rho \sin(\theta) = \phi$$

*Equation 45*

where $\theta$ is the phase difference between the injection and the oscillation, $\tau$ is the round-trip group delay, $\rho$ is the injection ratio, and $\phi$ is a measure of the detuning between oscillator and injected carrier.

The change of dependent variable:

$$x = \tan(\theta/2) \quad \Rightarrow \quad \sin(\theta) = \frac{2x}{1+x^2} \quad ; \quad \frac{dx}{dt} = \frac{1}{2}(1+x^2)\frac{d\theta}{dt}$$

*Equation 46*

transforms the Adler equation to:

$$2\tau \frac{dx}{dt} = \phi x^2 - 2\rho x + \phi$$

*Equation 47*

*For zero detuning* the equation simplifies to:

$$\tau \frac{dx}{dt} + \rho x = 0 \quad ; \quad \phi = 0$$

*Equation 48*

which has general solution:

$$x(t) = x(0)\exp(-(\rho/\tau)t) \quad ; \quad \phi = 0$$

*Equation 49*

The roots of the quadratic are:

$$x_\pm = \frac{1}{\phi}\left(\rho \pm \sqrt{\rho^2 - \phi^2}\right)$$

*Equation 50*

Either $\rho^2 > \phi^2$ and the roots are a reciprocal real pair $x_- = x_+^{-1}$ or $\rho^2 \leq \phi^2$ and the roots are a unimodular complex conjugate pair $x_- = x_+^*$  $|x_\pm| = 1$.

The reciprocal of the quadratic factors into two simple poles.

# Injection locked OEO multiscale phase dynamics

$$\frac{1}{x^2 - (2\rho/\phi)x + 1} = \frac{1}{(x_+ - x_-)}\left[\frac{1}{(x - x_+)} - \frac{1}{(x - x_-)}\right]$$

*Equation 51*

Hence:

$$\left[\frac{1}{(x - x_+)} - \frac{1}{(x - x_-)}\right]\tau\frac{dx}{dt} = \sqrt{\rho^2 - \phi^2}$$

*Equation 52*

or equivalently:

$$\tau\frac{d}{dt}\left[\ln\left(\frac{x - x_+}{x - x_-}\right)\right] = \sqrt{\rho^2 - \emptyset^2}$$

*Equation 53*

which has solution:

$$\frac{x(t) - x_+}{x(t) - x_-} = \frac{x(0) - x_+}{x(0) - x_-}\begin{cases} \exp\left(\sqrt{\rho^2 - \emptyset^2}\,\frac{t}{\tau}\right) & ; \rho^2 > \emptyset^2 \\ \exp\left(i\sqrt{\emptyset^2 - \rho^2}\,\frac{t}{\tau}\right) & ; \rho^2 \leq \emptyset^2 \end{cases}$$

*Equation 54*

where $x_\pm$ are given by Equation 50 and the solution $x$ must be real.

The case $\rho^2 > \emptyset^2$ corresponds to an asymptotically locked state. The roots $x_\pm$ are real and have the same sign, and the exponential factor is real and positive. Consequently, $(x(t) - x_+)/(x(t) - x_-)$ and $(x(0) - x_+)/(x(0) - x_-)$ must have the same sign. To retain the same sign $x(t)$ cannot cross either root yet it must tend to $x_-$ as $t \to \infty$ for an equilibrium state to exist. In the case where $x(0) \notin [x_-, x_+]$ this is only possible if $x(t)$ passes through the point at $\infty$ (the identification of $x = \pm\infty$ corresponds to the equivalence of $\theta = \pm\pi$ modulo $2\pi$).

The case $\rho^2 \leq \emptyset^2$ corresponds to a periodic unlocked state. The roots $x_\pm$ are complex conjugates and so $(x(t) - x_+)/(x(t) - x_-)$ and $(x(0) - x_+)/(x(0) - x_-)$ are unimodular which is consistent with the unimodular exponential factor.

*For the asymptotically locked state ($\rho^2 > \emptyset^2$):*

Set:

$$z = \exp(\gamma t + \chi) \quad ; \quad \gamma = \frac{1}{2}\frac{1}{\tau}\sqrt{\rho^2 - \emptyset^2} \quad ; \quad \chi = \frac{1}{2}\ln\left|\frac{x(0) - x_+}{x(0) - x_-}\right|$$

*Equation 55*

Hence:

# Injection locked OEO multiscale phase dynamics

$$\frac{x - x_+}{x - x_-} = z^2 \quad ; \quad \frac{x(0) - x_+}{x(0) - x_-} > 0$$

$$\Rightarrow$$

$$x = \frac{1}{2}(x_+ + x_-) - \left[\frac{z + z^{-1}}{z - z^{-1}}\right]\frac{1}{2}(x_+ - x_-)$$

*Equation 56*

Explicitly:

$$x(t) = \frac{1}{\emptyset}\rho - \frac{1}{\emptyset}\sqrt{\rho^2 - \emptyset^2}\coth(\gamma t + \chi)$$

*Equation 57*

Note that the condition $(x(0) - x_+)/(x(0) - x_-) > 0$ implies that $x(0) \notin [x_-, x_+]$ so if $x(0) > x_+$ then $x(t)$ must pass through $\infty$ at positive time. The necessary singular behaviour is provided by the hyperbolic cotangent function.

Or:

$$\frac{x - x_+}{x - x_-} = (iz)^2 \quad ; \quad \frac{x(0) - x_+}{x(0) - x_-} < 0$$

$$\Rightarrow$$

$$x = \frac{1}{2}(x_+ + x_-) - \left[\frac{z - z^{-1}}{z + z^{-1}}\right]\frac{1}{2}(x_+ - x_-)$$

*Equation 58*

Explicitly:

$$x(t) = \frac{1}{\emptyset}\rho - \frac{1}{\emptyset}\sqrt{\rho^2 - \emptyset^2}\tanh(\gamma t + \chi)$$

*Equation 59*

Note that the condition $(x(0) - x_+)/(x(0) - x_-) < 0$ implies that $x(0) \in [x_-, x_+]$ so $x(t)$ can reach $x_-$ without crossing $x_+$ or passing through $\infty$. Consequently, the necessary regular behaviour is provided by the hyperbolic tangent function.

Both the hyperbolic tangent and hyperbolic cotangent functions are asymptotic to unity as $t \to \infty$ verifying that $x \to x_-$.

*For the unlocked state ($\emptyset^2 > \rho^2$):*

Set:

$$z = \exp(i(\varpi t + \xi)) \quad ; \quad \varpi = \frac{1}{2}\frac{1}{\tau}\sqrt{\emptyset^2 - \rho^2} \quad ; \quad \xi = \frac{1}{2}\arg\left(\frac{x(0) - x_+}{x(0) - x_-}\right)$$

*Equation 60*

Hence:

# Injection locked OEO multiscale phase dynamics

$$\frac{x - x_+}{x - x_-} = z^2$$
$$\Rightarrow$$
$$x = \frac{1}{2}(x_+ + x_-) - \left[\frac{z + z^{-1}}{z - z^{-1}}\right]\frac{1}{2}(x_+ - x_-)$$

*Equation 61*

$$x(t) = \frac{1}{\phi}\rho - \frac{1}{\phi}\sqrt{\emptyset^2 - \rho^2}\cot(\varpi t + \xi)$$

*Equation 62*

The phase or the complex envelope may be recovered from the solutions for $x$ using:

$$\theta = 2\tan^{-1}(x)$$

*Equation 63*

or:

$$\exp(i\theta) = \frac{1 + ix}{1 - ix}$$

*Equation 64*

as convenient.

The solutions in terms of hyperbolic tangent and cotangent functions agree with those presented by Adler [11] accounting for differences in notation and Adler's use of the identity $\cot(\vartheta) = -\tan(\vartheta - \pi/2)$. The solution in terms of the hyperbolic cotangent appears to have been missed. It is however essential to the analytic solution of the two-time reduced phase model of injection locking because every possible initial condition can occur.

## Numerical considerations

The analytic solutions and their limits are well defined. Nevertheless, singularities inherent to the change of variable raise issues for numerical evaluation. With care these singularities may be removed.

The principal results in the preceding are:

$$x(t) = \exp(-\rho\, t/\tau)\quad ;\quad \phi = 0$$

$$\frac{x(t) - x_+}{x(t) - x_-} = \frac{x(0) - x_+}{x(0) - x_-}\begin{cases}\exp\left(\sqrt{\rho^2 - \emptyset^2}\ t/\tau\right) & ;\ \rho^2 > \emptyset^2 \\ \exp\left(i\sqrt{\emptyset^2 - \rho^2}\ t/\tau\right) & ;\ \rho^2 \le \emptyset^2\end{cases}$$

$$x_\pm = \frac{1}{\phi}\left(\rho \pm \sqrt{\rho^2 - \phi^2}\right)$$

*Equation 65*

# Injection locked OEO multiscale phase dynamics

These expressions define a unique function $x$ with initial value $x(0)$ for all time provided the roots $x_\pm$ are distinct $x_+ \neq x_-$. However, when the detuning is at the edge of the locking range where $|\phi| = \rho$, the roots are degenerate $x_+ = x_- = x_*$. It follows that:

$$\frac{x(t) - x_*}{x(t) - x_*} = \frac{x(0) - x_*}{x(0) - x_*}$$

*Equation 66*

The two sides of the equation are *independently* equal for any $x_*$ and any $x(0)$. Consequently, $x_*$ is not uniquely defined and its initial value can disagree with the initial value parameter. The resolution is to consider neighboring solutions $x$ for which $|\phi| \neq \rho$ but $|\phi| \sim \rho$ and consider the limit $|\phi| \to \rho$ from both above and below. The neighbouring solutions are uniquely defined, continuous, and agree with the initial value parameter. The only consistent choice of $x$ in the limit is:

$$x(t) = x(0) \quad ; \quad |\phi| = \rho$$

*Equation 67*

which implies:

$$\theta(t) = \theta(0) \quad ; \quad |\phi| = \rho$$

*Equation 68*

The singularity introduced by the initial condition:

$$x(0) = \tan(\theta(0)/2)$$

*Equation 69*

is resolved by use of the equality:

$$\frac{\phi x(0) - \phi x_+}{\phi x(0) - \phi x_-} = \frac{\phi \sin(\theta(0)/2) - \left(\rho + \sqrt{\rho^2 - \phi^2}\right)\cos(\theta(0)/2)}{\phi \sin(\theta(0)/2) - \left(\rho - \sqrt{\rho^2 - \phi^2}\right)\cos(\theta(0)/2)}$$

*Equation 70*

It follows that a better-behaved formulation of the solution for non-zero detuning is:

*For the asymptotically locked state ($\rho > |\emptyset|$)*

Set:

$$a = \phi \sin(\theta(0)/2) - \left(\rho + \sqrt{\rho^2 - \phi^2}\right)\cos(\theta(0)/2)$$
$$b = \phi \sin(\theta(0)/2) - \left(\rho - \sqrt{\rho^2 - \phi^2}\right)\cos(\theta(0)/2)$$
$$\chi = (1/2)\ln|a/b|$$
$$\gamma = (1/2)\frac{1}{\tau}\sqrt{\rho^2 - \emptyset^2}$$

*Equation 71*

Then:

# Injection locked OEO multiscale phase dynamics

$$\phi x(t) = \rho - \sqrt{\rho^2 - \emptyset^2} \coth(\gamma t + \chi) \quad ; \quad a/b > 0$$

*Equation 72*

or:

$$\phi x(t) = \rho - \sqrt{\rho^2 - \emptyset^2} \tanh(\gamma t + \chi) \quad ; \quad a/b < 0$$

*Equation 73*

It is possible for $a$ or $b$ to be zero and hence $\chi = \pm\infty$ but this is not an issue as MATLAB™ returns $\coth(\pm\infty) = \pm 1$ and $\tanh(\pm\infty) = \pm 1$. The case where both $a$ and $b$ are zero is excluded by the exception of the case of degenerate roots.

*For the unlocked state ($\rho < |\emptyset|$):*

Set:

$$a = \phi \sin(\theta(0)/2) - \left(\rho + i\sqrt{\phi^2 - \rho^2}\right) \cos(\theta(0)/2)$$
$$b = \phi \sin(\theta(0)/2) - \left(\rho - i\sqrt{\phi^2 - \rho^2}\right) \cos(\theta(0)/2)$$
$$\xi = (1/2) \arg(a/b) = \arg(a)$$
$$\varpi = (1/2)\frac{1}{\tau}\sqrt{\emptyset^2 - \rho^2}$$

*Equation 74*

Then:

$$\phi x(t) = \rho - \sqrt{\emptyset^2 - \rho^2} \cot(\varpi t + \xi)$$

*Equation 75*

*For either state:*

$$\exp(i\theta) = \frac{\phi + i\phi x}{\phi - i\phi x}$$

*Equation 76*

The parameter $\tau$ only enters as units of the time variable.